# Multiple Spawning with "Optimal" Basis Set Expansion

Sandy Yang, Joshua D. Coe, Benjamin Kaduk, and Todd J. Martínez

*Department of Chemistry, Beckman Institute, and Frederick Seitz Materials Research Laboratory, University of Illinois at Urbana-Champaign, 600 S. Mathews Ave., Urbana, IL 61801*

The Full Multiple Spawning (FMS) method is designed to simulate quantum dynamics in the multi-state electronic problem. The FMS nuclear wavefunction is represented in a basis of coupled, frozen Gaussians, and the "spawning" procedure prescribes a means of adaptively increasing the size of the basis in order to capture population transfer between electronic states. "Parent" trajectories create "children" when passing through regions of significant nonadiabatic coupling. In order to converge branching ratios without allowing the basis to reach an impractical size, population transfer at individual spawning events should be made as effective as possible. Herein we detail a new algorithm for specifying the initial conditions of freshly spawned basis functions, one that minimizes the number of spawns needed for convergence by maximizing the efficiency of individual spawning events. Optimization is achieved by maximizing the coupling between parent and child trajectories, as a function of child position and momentum, at the point of spawning. The method is tested with a two-state, one-mode avoided crossing model and a two-state, two-mode conical intersection model.

## I. INTRODUCTION

Conical intersections,[1] molecular configurations at which multiple electronic states become degenerate, have played an important role in photochemical reaction mechanisms[2-4] for some time.[3,5] They have come also to figure prominently in accounts of photodissociation,[4,6] photoisomerization,[4] ion fragmentation,[7] energy transfer,[8] and the basic chemistry underlying vision[9]. Direct simulation of dynamics in the neighborhood of an intersection requires methods able to describe multiple wavepackets evolving on different electronic surfaces, as well as the coherences connecting them. The full multiple spawning (FMS) method[10-13] meets all of these requirements, and has been successfully applied to photoisomerization in various hydrocarbons,[4,14] excited-state proton transfer in salicylic acid derivatives,[15,16] energy transfer in phenylacetylene dendrimers,[17] and photochemistry in DNA bases.[18,19] In addition, it is relatively straightforward to recast the results of an FMS simulation into forms directly comparable to time-resolved photoionization[18,20] and fluorescence upconversion[15,18,21] signals.

The total FMS wavefunction is built on a Born-Huang sum of products, the electronic and nuclear components of which are separately projected onto suitable function spaces. Standard basis sets (such as 6-31G*) typically are used to represent the electronic components, whereas the nuclear portions are expanded as coherent superpositions of frozen Gaussians. The nuclear basis set is adaptively enlarged through "spawning", a procedure by which an existing basis function (the "parent"), evolving on a single electronic surface, creates new basis functions ("children") on electronic surfaces to which it becomes nonadiabatically coupled in the course of its evolution. Because the reallocation of population between new and old basis functions is dictated by the Schrodinger equation (*vide infra*), not the spawning procedure itself, there is considerable freedom in choosing initial conditions for freshly spawned basis functions. To date, there have been two main approaches. The first passes the position vector of the parent on to the child, then adjusts the child's momentum along the direction of the nonadiabatic coupling in order to conserve energy. This method is most similar to standard surface hopping, and may lead to frustrated spawns.[22,23] The second procedure combines the momentum jump with a quench to the energy shell when spawns are frustrated by energy non-conservation. There are at least two ways of choosing the quench direction: steepest descent or descent along the nonadiabatic coupling vector. The right choice is surely ambiguous, but some of the relevant considerations will be discussed further below. Unfortunately, quenching procedures bias against actually *populating* new basis functions, because they displace centroids in both position and momentum spaces. This, in turn, lowers the maximum coupling between parent and child, promoting less population transfer between the two.

Because one of the primary aims of FMS is accurate prediction of population branching ratios, its overall efficiency is dictated largely by the effectiveness of individual spawning events in facilitating population transfer. It is worthwhile to consider, then, to what extent the initial conditions of newly spawned basis functions may be tuned in order to maximize this efficiency. The principal innovation of the present work is a third way of defining the initial conditions, one that we have dubbed "optimal" spawning. The basic concern is to minimize the number of basis functions required for convergence of branching ratios by maximizing the population transfer efficiency of individual spawning events. This is achieved by maximizing the matrix element coupling parent and child basis functions, as a function of child positions and momenta, at the point of spawning. In the process, all of the confusion surrounding descent directions disappears. As well as improving the efficiency of FMS simulations, the new procedure provides insight into a phase space picture of nonadiabatic phenomena.

Section 2 covers the FMS ansatz and equations of motion. Section 3 focuses specifically on the spawning mechanism, and the three approaches alluded to above are discussed in detail. This section forms the heart of the paper. As discussed there, the possibility of multiple local minima in the magnitude of the coupling between parent and child trajectories may lead to more than one optimally spawned basis function in a single event. The fourth section provides numerical comparison of wavefunctions and transmission coefficients generated by the three spawning methods. For testing purposes we have employed a one-dimensional, two-state avoided crossing model and a two-dimensional, two-state conical intersection model. For the latter model, this work presents the first rigorous comparison of FMS with numerically exact quantum dynamics. The final section concludes.

## II. FMS EQUATIONS OF MOTION

In *ab initio* multiple spawning (AIMS) [14], both the electrons and the nuclei are treated quantum mechanically and on a consistent basis. This consistency is difficult to achieve, at least numerically, because the locality of quantum chemistry conflicts with the global character of the nuclear Schrödinger equation. The focus of this paper is FMS (the core of AIMS), which incorporates quantum effects in the nuclear dynamics while remaining compatible with conventional quantum chemistry. This is achieved by localizing the nuclear Schrödinger equation through intelligent use of an adaptive, time-dependent basis set of frozen Gaussians, of a form pioneered by Heller and coworkers[24-27]. However, unlike the original frozen Gaussian approximation (FGA), FMS accounts fully for the nonorthogonal nature of the Gaussian basis. The multiconfigurational total wavefunction is written as

$$\psi = \sum_I \chi_I(\mathbf{R};t)\phi_I(\mathbf{r};\mathbf{R}) \tag{1}$$

The subscript $I$ indexes the electronic state, and vectors $\mathbf{r}$ and $\mathbf{R}$ denote the electronic and nuclear coordinates, respectively. Vectors and matrices are marked with boldface type. The electronic wavefunction $\phi_I(\mathbf{r};\mathbf{R})$ is an eigenfunction of the clamped-nucleus Hamiltonian $\hat{\mathbf{H}}_{el}(\mathbf{r};\mathbf{R})$, obtained by setting the nuclear kinetic energy in the total molecular Hamiltonian to zero. The semicolons in $\phi_I(\mathbf{r};\mathbf{R})$ and $\hat{\mathbf{H}}_{el}(\mathbf{r};\mathbf{R})$ indicate a parametric dependence on $\mathbf{R}$. $\chi_I(\mathbf{R};t)$ is the time-dependent nuclear wavefunction associated with the electronic state $I$, and the set of $\chi_I(\mathbf{R};t)$ act as expansion coefficients in the Born-Oppenheimer representation.[28] Frozen Gaussians are used to represent the nuclear wavefunction[24,26,29-32] for each electronic state,

$$\chi_I(\mathbf{R};t) = \sum_{m=1}^{N_I(t)} c_m^I(t)\chi_m^I\left(\mathbf{R};\overline{\mathbf{R}}_m,\overline{\mathbf{P}}_m,\gamma_m,\alpha_m\right), \tag{2}$$

where $N_I(t)$ is the number of nuclear basis functions on electronic state $I$ at time $t$. The value of $N_I(t)$ may change during the propagation, *e.g.*, when a new trajectory is spawned. The $c_m^I$ are complex, time-dependent amplitudes paired with the $\chi_m^I$, which we will interchangeably reference as "basis functions" or "trajectories." The full quantum dynamics is described by the bundle of trajectories embodied in sums (1) and (2). Each individual nuclear basis function $\chi_m^I$ can be written as a multidimensional product of one-dimensional Gaussians $\chi_{\rho_m}^I$,

$$\chi_m^I\left(\mathbf{R}; \overline{\mathbf{R}}_m^I, \overline{\mathbf{P}}_m^I, \gamma_m^I, \alpha_m^I\right) = e^{i\gamma_m^I(t)} \prod_{\rho=1}^{F} \chi_{\rho_m}^I\left(R; \overline{R}_{\rho_m}^I, \overline{P}_{\rho_m}^I, \alpha_{\rho_m}^I\right), \qquad (3)$$

where $F$ is the total number of system degrees of freedom. Each one-dimensional frozen Gaussian is given as

$$\chi_{\rho_m}^I\left(R; \overline{R}_{\rho_m}^I, \overline{P}_{\rho_m}^I, \alpha_{\rho_m}^I\right) = \left(\frac{2\alpha_{\rho_m}^I}{\pi}\right)^{1/4} e^{-\alpha_{\rho_m}^I\left(R-\overline{R}_{\rho_m}^I\right)^2 + i\overline{P}_{\rho_m}^I\left(R-\overline{R}_{\rho_m}^I\right)} \qquad (4)$$

The Gaussians are parameterized by centroid position $\overline{R}_{\rho_m}^I$, momentum $\overline{P}_{\rho_m}^I$, width $\alpha_{\rho_m}^I$, and phase $\gamma_m^I$. Note that $\gamma_m^I$ is absorbed into the multidimensional $\chi_m^I$, such that a single nuclear phase factor $\gamma_m^I$ is associated with each product of Gaussians $\chi_{\rho_m}^I$. Numerical considerations encourage the use of a fixed, time-independent width parameter $\alpha_{\rho_m}^I$. In the special case of a harmonic potential, the natural choice of width is determined completely by mass and frequency.[24,27] For an arbitrary potential, the optimal choice of widths is unclear. In general, however, simulation results are insensitive to the particular values chosen for $\alpha_{\rho_m}^I$, so long as they fall within a fairly broad range.[13] An upcoming work will provide further detail regarding the choice of widths for atoms in various systems. Gaussian widths $\alpha_{\rho_m}^I$ are chosen without reference to electronic state $I$ or nuclear basis function $j$, *i.e.*, $\alpha_{\rho_j}^I \to \alpha_\rho \ \forall\ I, j$. Due to the insensitivity of the results, $\alpha_\rho$ can be assigned any reasonable value so long as stability and rate of convergence are not sacrificed.

With most conventional *ab initio* molecular dynamics methods, special attention must be paid to the interface of the quantum mechanical treatment of the electrons and that of the nuclei. This results from the tension between the locality of quantum chemistry and the global character of the nuclear Schrödinger equation. One would like a method that retained classical mechanics (complete locality) in one limit while capturing quantum effects fully and exactly in another. FMS introduces locality in the nuclear dynamics by employing frozen Gaussian basis functions and choosing a classical propagator for the centroid positions and momenta. Many research groups have investigated the use of Gaussian wavepackets in both semiclassical[24,26,29,30] and quantum mechanical[31,32] dynamics. The use of FGA in FMS bears most resemblance to work performed by Metiu and coworkers.[30,31] Their work yielded quantitative accuracy through employment of a variational principle. This approach is less desirable for FMS, because it is incapable of reduction to the classical limit. Another, more practical, reason is that variational approaches are more expensive computationally, requiring more matrix elements and their derivatives, and thus conflicting with the locality of quantum chemistry. Hamilton's equations of motion for parameters $\bar{R}^I_{\rho_m}$ and $\bar{P}^I_{\rho_m}$ are given by

$$\frac{\partial \bar{R}^I_{\rho_m}}{\partial t} = \frac{\bar{P}^I_{\rho_m}}{m_\rho}$$

$$\frac{\partial \bar{P}^I_{\rho_m}}{\partial t} = -\frac{\partial V^I(\mathbf{R})}{\partial R}\bigg|_{\bar{R}^I_{\rho_m}} \quad (5)$$

Each basis function centroid evolves as a classical trajectory on a single electronic potential energy surface. Nuclear phase $\gamma^I_m$ advances by the semiclassical prescription

$$\frac{\partial \gamma^I_m}{\partial t} = \sum_{\rho=1}^{F} \frac{\left(\bar{P}^I_{\rho_m}\right)^2 - 2\alpha^I_{\rho_m}}{2 M_\rho} - V^I\left(\bar{\mathbf{R}}^I_m\right). \quad (6)$$

Note that the phase factor $\gamma^I_m$ is redundant with the complex coefficient $c^I_m$; therefore, we arbitrarily choose its equation of motion as the one obtained in the local harmonic approximation[12,24]. $V^I(\mathbf{R})$ represents the potential energy at position **R** on electronic state *I*, and $M_\rho$ is the mass of the $\rho$-th coordinate.

Although the Gaussian basis function parameters evolve classically, fully quantum dynamics is preserved by solving the Schrödinger equation for the coefficients in (2). The exact quantal amplitudes $c_m^I$ obey the following equation of motion

$$\frac{d\mathbf{C}^I}{dt} = -i\left(\mathbf{S}^{II}\right)^{-1}\left\{\left(\mathbf{H}^{II} - i\dot{\mathbf{S}}^{II}\right)\mathbf{C}^I + \sum_{J \neq I} \mathbf{H}^{IJ}\mathbf{C}^J\right\} \tag{7}$$

where $\mathbf{C}^I = \{c_m^I; m = 1, 2, ..., N\}$ is composed of the complex coefficients $c_m^I$ for each trajectory. The matrices $\mathbf{H}$ and $\mathbf{S}$ are defined by integration over electronic coordinate $\mathbf{r}$,

$$\mathbf{H}^{IJ} \equiv \left\langle \phi_I(\mathbf{r};\mathbf{R}) \left| \hat{\mathbf{H}} \right| \phi_J(\mathbf{r};\mathbf{R}) \right\rangle_\mathbf{r} \tag{8}$$

and

$$\mathbf{S}^{IJ} \equiv \left\langle \phi_I(\mathbf{r};\mathbf{R}) \left| \hat{\mathbf{1}} \right| \phi_J(\mathbf{r};\mathbf{R}) \right\rangle_\mathbf{r} \tag{9}$$

Because the electronic states are orthogonal by construction (in the adiabatic representation, at least), $\mathbf{S}^{IJ}$ vanishes for all $I \neq J$, i.e., $\mathbf{S}$ acts as an identity operator. However, $\mathbf{S}$ is not diagonal in the nuclear indices because the frozen Gaussians are nonorthogonal. The matrix elements of $\mathbf{S}$ are given by

$$S_{mn}^{II} = \left\langle \chi_m^I\left(\mathbf{R};\overline{\mathbf{R}}_m^I,\overline{\mathbf{P}}_m^I,\gamma_m^I\right) \middle| \chi_n^I\left(\mathbf{R};\overline{\mathbf{R}}_n^I,\overline{\mathbf{P}}_n^I,\gamma_n^I\right) \right\rangle_\mathbf{R}, \tag{10}$$

where the integration is performed over the (dummy) nuclear coordinates $\mathbf{R}$. Note that $\mathbf{S}^{-1}$ in (7) accounts fully for the nonorthogonal character of the Gaussian basis set. Because the basis functions $\chi_m^I$ are time-dependent, propagation of the quantal amplitudes $c_m^I$ requires also the time derivative of the overlap matrix, defined by

$$\dot{S}_{mn}^{II} = \left\langle \chi_m^I \left| \frac{\partial}{\partial t} \right| \chi_n^I \right\rangle. \tag{11}$$

On the other hand, the off-diagonal elements of $\hat{\mathbf{H}}$ are nonzero in both electronic and nuclear indices

$$\mathbf{H}_{mn}^{IJ} \equiv \left\langle \phi_I(\mathbf{r};\mathbf{R})\chi_m^I\left(\mathbf{R};\overline{\mathbf{R}}_m^I,\overline{\mathbf{P}}_m^I\right) \middle| \hat{\mathbf{T}}_\mathbf{R} + \hat{\mathbf{H}}_{el} \middle| \phi_J(\mathbf{r};\mathbf{R})\chi_n^J\left(\mathbf{R};\overline{\mathbf{R}}_n^J,\overline{\mathbf{P}}_n^J\right) \right\rangle_{\mathbf{r},\mathbf{R}}. \tag{12}$$

$\hat{\mathbf{T}}_\mathbf{R}$ denotes the nuclear kinetic energy operator, and $\hat{\mathbf{H}}_{el}$ the clamped-nucleus Hamiltonian. This matrix element may be further decomposed as

$$\left(\hat{\mathbf{T}}_\mathbf{R}\right)_{mn}^{IJ} \equiv \delta_{IJ} \left\langle \chi_m^I\left(\mathbf{R};\overline{\mathbf{R}}_m^I,\overline{\mathbf{P}}_m^I,\gamma_m^I\right) \middle| \sum_{\rho=1}^{3N} \frac{-1}{2m_\rho}\frac{\partial^2}{\partial R_\rho^2} \middle| \chi_n^J\left(\mathbf{R};\overline{\mathbf{R}}_n^J,\overline{\mathbf{P}}_n^J,\gamma_n^J\right) \right\rangle_\mathbf{R}$$

$$+ \left\langle \chi_m^I\left(\mathbf{R};\overline{\mathbf{R}}_m^I,\overline{\mathbf{P}}_m^I,\gamma_m^I\right) \middle| 2D^{IJ} + G^{IJ} \middle| \chi_n^J\left(\mathbf{R};\overline{\mathbf{R}}_n^J,\overline{\mathbf{P}}_n^J,\gamma_n^J\right) \right\rangle_\mathbf{R}, \quad (13)$$

with

$$\left(\hat{\mathbf{H}}_{el}\right)_{mn}^{IJ} \equiv \left\langle \chi_m^I\left(\mathbf{R};\overline{\mathbf{R}}_m^I,\overline{\mathbf{P}}_m^I,\gamma_m^I\right) \middle| V^{IJ}(\mathbf{R}) \middle| \chi_n^J\left(\mathbf{R};\overline{\mathbf{R}}_n^J,\overline{\mathbf{P}}_n^J,\gamma_n^J\right) \right\rangle_\mathbf{R}. \quad (14)$$

The first term in the nuclear kinetic matrix element (13) can be evaluated analytically due to the Gaussian form of the nuclear wavefunctions $\chi_m^I$. The second and third terms are defined by

$$D^{IJ} = \sum_{\rho=1}^{3N} \frac{-d_\rho^{IJ}}{2M_\rho} \frac{\partial}{\partial R_\rho} \quad (15)$$

$$G^{IJ} = \sum_{\rho=1}^{3N} \frac{-1}{2M_\rho} \left\langle \phi_I(\mathbf{r};\mathbf{R}) \middle| \frac{\partial^2}{\partial R_\rho^2} \middle| \phi_J(\mathbf{r};\mathbf{R}) \right\rangle_\mathbf{r}, \quad (16)$$

in which $d_\rho^{IJ}$ is an element of the nonadiabatic coupling vector $\mathbf{d}^{IJ}$,

$$\mathbf{d}^{IJ} \equiv \left\langle \phi_I(\mathbf{r};\mathbf{R}) \middle| \frac{\partial}{\partial \mathbf{R}} \middle| \phi_J(\mathbf{r};\mathbf{R}) \right\rangle_\mathbf{r}. \quad (17)$$

$\mathbf{d}^{IJ}$ is also referred to as the derivative coupling vector. $G^{IJ}$ is related to the scalar coupling $\equiv \left\langle \phi_I(\mathbf{r};\mathbf{R}) \middle| \nabla^2 \middle| \phi_J(\mathbf{r};\mathbf{R}) \right\rangle$, where $\nabla \equiv \frac{\partial}{\partial \mathbf{R}}$. By virtue of the Hellmann-Feynman theorem, the off-diagonal elements of $\mathbf{d}^{IJ}$ can be written in the form

$$\mathbf{d}^{IJ} = \frac{\left\langle \phi_I(\mathbf{r};\mathbf{R}) \middle| (\nabla \hat{\mathbf{H}}_{el}) \middle| \phi_J(\mathbf{r};\mathbf{R}) \right\rangle}{V^{JJ}(\mathbf{R}) - V^{II}(\mathbf{R})} \quad (18)$$

One finds, therefore, that the nonadiabatic coupling vector exhibits an inverse dependence on the energy gap separating the two potential surfaces. As the gap narrows to zero, the mass factor $M_\rho$ becomes negligible and the coupling of nuclear motion on the two electronic states becomes singular.

The precise form of the coupling depends on the representation of the electronic wavefunction. In the *adiabatic* representation, the wavefunctions $\phi_I(\mathbf{r};\mathbf{R})$ are the eigenfunctions of the clamped-nucleus Hamiltonian $\hat{H}_{el}(\mathbf{r};\mathbf{R})$, with eigenvalues $V^{II}(\mathbf{R})$. The electronic component of the full Hamiltonian in (12), $\hat{\mathbf{H}}_{el}$, is block diagonal, the blocks corresponding to different electronic states. The off-diagonal elements of the nuclear kinetic operator $\hat{\mathbf{T}}_\mathbf{R}$, $D^{IJ}$ and $G^{IJ}$, do not vanish, and thereby constitute the nonadiabatic coupling. Therefore, the matrix elements of the Hamiltonian operator in the *adiabatic* representation are given by

$$H_{mn}^{IJ} = \delta_{IJ} \cdot \left\langle \chi_m^I\left(\mathbf{R}; \overline{\mathbf{R}}_m^I, \overline{\mathbf{P}}_m^I, \gamma_m^I\right) \middle| V^{IJ}(\mathbf{R}) \middle| \chi_n^J\left(\mathbf{R}; \overline{\mathbf{R}}_n^J, \overline{\mathbf{P}}_n^J, \gamma_n^J\right) \right\rangle_\mathbf{R}$$

$$+ \delta_{IJ} \cdot \left\langle \chi_m^I\left(\mathbf{R}; \overline{\mathbf{R}}_m^I, \overline{\mathbf{P}}_m^I, \gamma_m^I\right) \middle| \sum_{\rho=1}^{3N} \frac{-1}{2M_\rho} \frac{\partial^2}{\partial R_\rho^2} \middle| \chi_n^J\left(\mathbf{R}; \overline{\mathbf{R}}_n^J, \overline{\mathbf{P}}_n^J, \gamma_n^J\right) \right\rangle_\mathbf{R}$$

$$+ \left\langle \chi_m^I\left(\mathbf{R}; \overline{\mathbf{R}}_m^I, \overline{\mathbf{P}}_m^I, \gamma_m^I\right) \middle| 2D^{IJ} + G^{IJ} \middle| \chi_n^J\left(\mathbf{R}; \overline{\mathbf{R}}_n^J, \overline{\mathbf{P}}_n^J, \gamma_n^J\right) \right\rangle_\mathbf{R} \quad (19)$$

When the molecular Schrödinger equation is recast as a set of coupled eigenvalue equations depending on nuclear coordinates $\mathbf{R}$ only, off-diagonal elements in these equations arise from the nuclear kinetic energy operator. The adiabatic representation is converted to the *diabatic* representation by a unitary transformation. The diabatic representation replaces off-diagonal kinetic energy terms by off-diagonal potential energy terms. Changes in electronic character due to nuclear perturbations are minimized, giving rise to smoother potential energy surfaces. In particular, there are no singularities at surface crossings. In the diabatic representation, the first term in Eq. (19) is no longer diagonal, and the off-diagonal matrix elements $V^{IJ}(\mathbf{R})$ are responsible for coupling the electronic states. Conversely, the off-diagonal elements of $D^{IJ}$ and $G^{IJ}$ vanish. Eq. (19) is replaced by

$$H_{mn}^{IJ} = \left\langle \chi_m^I\left(\mathbf{R};\overline{\mathbf{R}}_m^I,\overline{\mathbf{P}}_m^I,\gamma_m^I\right)\middle|V^{IJ}(\mathbf{R})\middle|\chi_n^J\left(\mathbf{R};\overline{\mathbf{R}}_n^J,\overline{\mathbf{P}}_n^J,\gamma_n^J\right)\right\rangle_\mathbf{R}$$
$$+\delta_{IJ}\cdot\left\langle \chi_m^I\left(\mathbf{R};\overline{\mathbf{R}}_m^I,\overline{\mathbf{P}}_m^I,\gamma_m^I\right)\middle|\sum_{\rho=1}^{3N}\frac{-1}{2M_\rho}\frac{\partial^2}{\partial R_\rho^2}\middle|\chi_n^J\left(\mathbf{R};\overline{\mathbf{R}}_n^J,\overline{\mathbf{P}}_n^J,\gamma_n^J\right)\right\rangle_\mathbf{R}. \quad (20)$$

## III. BASIS SET EXPANSION

### A. General spawning algorithm

The equation of motion (7) leads to the full multiple spawning method, which was developed to describe nonadiabatic transitions in the multi-state electronic problem. As the core of FMS, the spawning procedure for adding basis functions during nonadiabatic events is the key to the method's accuracy and efficiency. The spawning technique must provide numerical convergence while keeping the basis size manageable. This is particularly important in the multidimensional case. Because nonadiabatic events usually are marked by strong nonadiabatic coupling, conventional FMS allows basis functions on one electronic state to spawn new trajectories on another electronic state only when they enter a region of nonadiabatic coupling. By this means, basis set growth due to spawning new trajectories is controlled even while maintaing reasonable wavefunction accuracy.

The spawning procedure dictates when and how to spawn. Spawning regions are demarcated by threshold values of the effective nonadiabatic coupling

$$\Lambda^{IJ}(\mathbf{R}) = \begin{cases} \left|\dfrac{V^{IJ}(\mathbf{R})}{V^{II}(\mathbf{R})-V^{JJ}(\mathbf{R})}\right| & \text{diabatic} \\ \left|\dot{\mathbf{R}}\cdot\mathbf{d}^{IJ}\right| & \text{adiabatic} \end{cases} \quad (21)$$

The nonadiabatic coupling vector $\mathbf{d}^{IJ}$ is defined by (17), and $\dot{\mathbf{R}}$ is the nuclear velocity vector. In the diabatic representation, the nonadiabatic coupling depends only on the nuclear coordinates. In the adiabatic representation, however, the equivalent expression depends on the nuclear velocity.[11] In both representations, basis functions are counted as being in a nonadiabatic region as soon as the effective coupling (21) exceeds some predetermined threshold; we denote this initial time $t_i$. Once a trajectory enters a spawning region, it is propagated until the effective nonadiabatic coupling falls below the spawning threshold; this time we label $t_f$. The difference $t_f$-$t_i$ is the crossing time. Use of a nonadiabatic coupling criterion helps minimize the number of spawning trials, and is crucial to the overall efficiency of the spawning procedure. New trajectories are spawned only when their overlaps with all preexisting trajectories are small, otherwise the spawned basis function will be redundant and the method's efficiency will suffer. Perhaps worse, creating new basis functions that strongly overlap with existing ones may lead to linear dependence due to the overcompleteness of the Gaussian set.[11,12]

Upon entering a spawning region, a trajectory is *forward-propagated* through to its end according to (5). Its evolution over the course of this period is uncoupled from that of the remaining bundle, *i.e.*, its complex amplitude is not propagated as prescribed by (7). The parent trajectory may spawn several new basis functions during this time, although the precise number is an input parameter. In the simplest case, this parameter is set to one and the new function is created at the point of maximum nonadiabatic coupling $t_1$. Summarizing the above description, a necessary condition for spawning is that a trajectory's effective coupling $\Lambda^{IJ}$ to another electronic state exceed a threshold value. In addition, the maximal overlap between the new trajectory and those in the existing bundle should fall below a preset threshold value. This additional constraint minimizes unnecessary growth of the basis set and helps avoid problems associated with linear dependence.

Once a new basis function has been spawned, assuming that its initial position and momentum have been assigned (by means not yet specified), both new and old trajectories are *backward-propagated* from the spawning point $t_1$ to the starting time $t_i$ of the effective coupling region. As in the case of the forward-propagation, the backward-propagation of the parent-child pair is uncoupled from that of the bundle as a whole. The new basis function is added to the bundle at $t_i$, at which point coupled propagation of all basis functions resumes according to (5)-(7).

**B. Nonadiabatic coupling and energy conservation**

As mentioned previously, there are two major ingredients to spawning: when and how to spawn. Having determined the "when" in the previous section, we now cover the "how" – namely, where in phase space to place the newly spawned trajectory. Energy conservation is an important consideration in this context; although FMS is based on exact solution of the Schrödinger equation, for numerical convenience it is the *classical* energy of the system that is actually conserved, in agreement with standard molecular dynamics (MD). It is worth noting, however, that quantum and classical energies are noticeably different only in regions where basis functions are coupled by off-diagonal matrix elements of the Hamiltonian. The issue becomes subtler when there are many initial basis functions having different energies and encountering extended coupling regions. However, for many-dimensional systems, any finite number of initial basis functions will be negligibly coupled in the long time limit due to exponential divergence in phase space[27,33]. Therefore, one can safely assume that the trajectory ensemble will behave asymptotically as an incoherent superposition of independent basis functions. In such a limit, it is reasonable to believe that the difference between classical and quantum energies will be roughly equal modulo the zero-point energy correction. This is why classical MD has proven so successful in simulating large chemical and biological systems.

The connection with classical mechanics suggests that one place the new trajectory on the classical energy shell of its parent. This conflicts, however, with the first-order perturbation result, which predicts child trajectories proportional to the product of their parent and the nonadiabatic coupling function. For example, if the nonadiabatic coupling is independent of $\mathbf{R}$, perturbation theory places the child trajectory at precisely the same phase point as its parent. In general, this way of initializing new basis functions does not conserve classical energy unless the parent lies exactly at the crossing seam (in the diabatic representation) or at a conical intersection (in the adiabatic). On the other hand, conservation of classical energy in the long-time limit is clearly desirable.[34] The state-to-state form of Fermi's golden rule gives the transition probability from state *i* to *j* as

$$\Gamma_{i \to j} = \frac{2\pi}{\hbar} \left| H_{ij} \right|^2 \delta\left(E_j - E_i\right) \quad (22)$$

The only allowed nonadiabatic transitions, asymptotically, are those that conserve energy. It is reasonable to expect, then, that the expectation value of the energy will interpolate smoothly between these two limits.

Consider a simplified case in which the FMS bundle begins as a single trajectory $m$ on electronic state $I$. Trajectory $m$ enters a strong nonadiabatic coupling region and spawns a new trajectory $n$ on electronic state $J$. Asymptotically, when both $m$ and $n$ have left the region of nonadiabatic coupling, there is no interference between the two and the total energy is

$$\langle \varphi(\mathbf{R};t)|\hat{\mathbf{H}}|\varphi(\mathbf{R};t)\rangle = |c_m^I|^2 \langle \chi_m^I(\mathbf{R};t)|\mathbf{H}^{II}|\chi_m^I(\mathbf{R};t)\rangle + |c_n^J|^2 \langle \chi_n^J(\mathbf{R};t)|\mathbf{H}^{JJ}|\chi_n^J(\mathbf{R};t)\rangle . \quad (23)$$

Norm and energy conservation imply equality of the two expectation values on the right hand side of Eq. (23). Classically, this equality implies that the average energy of the basis functions representing the wavepacket on each electronic state is the same. In addition, because each trajectory evolves according to the classical equations of motion (5), the classical energy of each trajectory,

$$\left(E_{CL}\right)_m^I = |c_m^I|^2 \left( V^{II}\left(\bar{R}_m^I\right) + \sum_{\rho=1}^{F} \frac{\left(\bar{P}_{\rho_m}^I\right)^2}{2 M_{\rho_m}} \right), \quad (24)$$

is conserved throughout the propagation. The combination of (5), (22), and (24) implies that parent and child basis functions should have the same classical energy at the time of spawning. This constraint leads to the following overcomplete set of equations:

$$\frac{\partial}{\partial \bar{\mathbf{P}}_\rho^J}\left|\langle \chi_m^I|H^{IJ}|\chi_n^J\rangle\right| = 0$$

$$\frac{\partial}{\partial \bar{\mathbf{R}}_\rho^J}\left|\langle \chi_m^I|H^{IJ}|\chi_n^J\rangle\right| = 0 \quad (25)$$

$$V^{II}\left(\bar{\mathbf{R}}_m^I\right) + \sum_{\rho=1}^{3N} \frac{\left(\bar{P}_{\rho_m}^I\right)^2}{2 M_\rho} = V^{JJ}\left(\bar{\mathbf{R}}_n^J\right) + \sum_{\rho=1}^{3N} \frac{\left(\bar{P}_{\rho_n}^J\right)^2}{2 M_\rho}$$

In general, (25) does not have a unique solution, and can be solved only in a least-squared sense. As the number of basis functions increases, however, and for finite widths in the Gaussian basis, the results are insensitive to the details of the solution. It is instructive to study the simplest case once again, a pair of one-dimensional diabats with $R$-independent coupling. Maximizing $\left|\langle \chi_m^I|\mathbf{H}^{IJ}|\chi_n^J\rangle\right|$ is then equivalent to maximizing the overlap integral $\left| \left\langle \chi_m^I \left( \mathbf{R}; \overline{\mathbf{R}}_m^I, \overline{\mathbf{P}}_m^I, \gamma_m^I \right) \middle| \chi_n^J \left( \mathbf{R}; \overline{\mathbf{R}}_n^J, \overline{\mathbf{P}}_n^J, \gamma_n^J \right) \right\rangle_\mathbf{R} \right|$. The overlap integral of two Gaussians contains arguments of the form $R_m^I - R_n^J$ and $P_m^I - P_n^J$, motivating two approximate solutions: *position-preserving* and *momentum-preserving* spawns. In these limits, either position or momentum is fixed while the conjugate variable is adjusted to equalize classical energy between parent and child trajectories. In one dimension, the position-preserving spawn is equivalent to the momentum-jump used in surface-hopping.[23,35,36] If the energy is low enough, it may be impossible for the position-preserving method to satisfy the energy constraint, whereas momentum-preserving spawns allow for tunneling.[12] This is a very special type of tunneling, however, dependent on the representation used for the electronic wavefunction. For adiabatic representations, the tunneling described by momentum-preserving spawns leads to intrastate spawning, although current implementations of FMS allow only interstate spawning. This point is discussed in more detail in the next section.

We now address one last issue regarding energy conservation. Off-diagonal elements of the Hamiltonian (coherences) contribute only to the quantum mechanical expression. Classically, the potential energy of each trajectory is equivalent to that of a particle placed at the Gaussian center $\overline{R}_i$, and the kinetic energy is given by (24). To evaluate the corresponding quantum mechanical energy, one needs the Gaussian integral for kinetic and potential energy operators, $\left\langle \chi_m^I \middle| \hat{\mathbf{T}} + \hat{\mathbf{V}} \middle| \chi_n^J \right\rangle$. The difference between the two expressions is only a small constant after trajectories have decoupled from one another, when off-diagonal matrix elements of $S_{ij} = \left\langle \chi_i \middle| \chi_j \right\rangle$ all are negligible. For example, the analytical expression for the one-dimensional quantum mechanical kinetic energy can be written as

$$\left( \mathbf{T}_{\mathrm{QM}} \right)_{mm}^{II} = \frac{1}{2M} \left\langle \chi_m^I \left| -\frac{d^2}{d\hat{x}^2} \right| \chi_m^I \right\rangle = \left( \mathbf{T}_{\mathrm{CL}} \right)_{mm}^{II} + \frac{\alpha}{2M} \qquad (26)$$

The difference between the quantum and classical kinetic energies is simply the zero point energy[33,37,38]. Potential energy analysis depends on the details of the potential model, but similar conclusions hold: differences between the two expressions are independent of $\bar{\mathbf{R}}_m^I$ and $\bar{\mathbf{P}}_m^I$.

**C. Spawning procedures**

*1. Momentum adjustment along the nonadiabatic coupling vector*

Given the demand that spawned basis functions have the same classical energy as their parent, initial conditions for the newly spawned trajectories can be determined by solving the overcomplete set of equations (25). The solution is, in general, non-unique, leading to various choices of additional constraint. It has been shown that momentum-preserving or position-preserving spawns represent two simple, approximate solutions. We start with the simplest solution of a position-preserving spawn paired with a pure momentum jump.

For a one-dimensional system, momentum jumps are clearly defined and position-preserving spawns lead to the same adjustment used in surface hopping[22,23]. This procedure has been justified semiclassically by Herman.[35] In practice, the momentum of a new trajectory is calculated as

$$\mathbf{P}_{new}^I = \mathbf{P}_{old}^J - D\hat{\mathbf{d}}^{IJ}, \qquad (27)$$

where $\mathbf{P}_{new}^I$ is the centroid momentum vector of the newly spawned child, $\mathbf{P}_{old}^J$ is that of the parent, and $\hat{\mathbf{d}}^{IJ}$ is a unit vector directed along the nonadiabatic coupling defined in (17),

$$\hat{\mathbf{d}}^{IJ} = \frac{\mathbf{d}^{IJ}}{\left|\mathbf{d}^{IJ}\right|}. \qquad (28)$$

$D$ is a scalar, the value of which is chosen such that the total kinetic energy of the parent is identical to that of the child.

Sometimes the surface to which a spawn should occur is classically inaccessible, in that removing the component of $\mathbf{P}_{old}^J$ along $\hat{\mathbf{d}}^{IJ}$ is not enough to compensate for the energy deficit. In surface hopping, such failures are known as frustrated hops. By allowing trajectories to branch, surface hopping enables trajectories to switch quantum states stochastically at any point. When a trial hop is frustrated, it is discarded and the trajectory remains on its original electronic state. Many have tried to tease out the full implications of frustrated hops,[39,40] and recently they were identified as essential to proper equilibrium.[39] Frustrated hops are the means by which surface hopping satisfies detailed balance, which is necessary to ensure accurate evolution in short time.

In FMS, however, frustrated spawns do not affect the maintenance of detailed balance. The Schrödinger equation, not the spawning algorithm, governs population transfer. The spawning procedure is important in that it allows nonadiabatic transitions in the first place, but spawning itself does not dictate the statistical balance of population among the various electronic surfaces. A trajectory may spawn new basis functions upon passage through a nonadiabatic coupling region, but then the Schrödinger equation is solved exactly in order to determine the time-dependent population associated with each trajectory. Therefore, one need not take into account the effect of frustrated spawns. The amount of population transferred to child trajectories in frustrated spawns typically is small, leading to less efficient but still accurate population transfer dynamics. If the kinetic energy along the nonadiabatic coupling vector is insufficient to conserve total energy when a trajectory attempts to spawn to a higher-lying electronic surface, the spawn simply is accepted and all kinetic energy along the nonadiabatic direction is removed. Detailed balance is satisfied automatically by solving the Schrödinger equation.

Even using the simplest pure momentum jump, FMS differs from standard surface hopping in its backward propagation of parent-child pairs. Various spawning algorithms may differ in their manner of defining initial conditions for newly spawned trajectories, but backward propagation is performed regardless. After each parent-child pair has been backward-propagated to the state at which the parent entered the nonadiabatic coupling region, the energies of the parent and child trajectories are used to determine whether the upcoming spawning event will be frustrated or not. This delay in the energy conservation test introduces a small probability of the system "tunneling" through the potential barrier by "borrowing" some energy during the back propagation.

*2. Momentum jump with quench to the energy shell*

As mentioned previously, position-preserving spawns can be frustrated when the surface to which they should occur is energetically inaccessible. In these cases, rescaling the component of momentum along the nonadiabatic coupling vector does not provide the kinetic energy needed to jump to the higher energy electronic state. One could simply discard the frustrated child trajectory, without affecting accuracy. This simple approach is not very effective, however, and many important spawning events might be skipped and FMS much less efficient.

There are alternatives, such as quenching to the energy shell. There are two obvious choices of descent direction, either of which performs reasonably well: the nonadiabatic coupling vector, or the negative of the gradient (steepest descent). If the nonadiabatic coupling vector $\hat{\mathbf{d}}^{IJ}$ is chosen, then the momentum-jump is followed by a position shift that minimizes the functional

$$E\left(\mathbf{R}_{\text{old}}^{I} + \gamma \cdot \hat{\mathbf{d}}^{IJ}\right) - E\left(\mathbf{R}_{\text{old}}^{I}\right). \tag{29}$$

$\mathbf{R}_{\text{old}}$ refers to the position vector of the parent trajectory, and $\gamma$ is computed to minimize (29). For steepest descent, one need only replace $\hat{\mathbf{d}}^{IJ}$ with the unit vector gradient and a change of sign. The gradient, in this case, is $\partial V^{II}(\mathbf{R})/\partial \mathbf{R}$, in which $I$ indexes the electronic state of the child trajectory. For one-dimensional problems, the two approaches are identical. In general, however, their relative accuracies vary with time and system, and the better choice is unclear.

The combination of momentum jump and steepest descent generally leads to spawned trajectories that do not preserve position or momentum of parent trajectories. This introduces a bias against fully populating newly spawned trajectories, because the maximum overlap between parent and child is poor. This diminished population transfer actually ensures detailed balance, but we will not explore this issue further here.

*3. Optimal spawning*

Previous sections surveyed pure position and pure momentum jumps, but the key to spawning optimally lies in pinpointing the ideal blend of the two. Heller and coworkers[38,41] noted the importance of hybrid jumps, but (to our knowledge) a method acting on this principle has thus far failed to appear. The proper mix is found by minimizing the functional,

$$\lambda E_{\text{diff}} - V_{\text{pc}}^{IJ}, \tag{30}$$

where the individual terms are defined by

$$E_{\text{diff}} = \left| E\left(\chi_{\text{parent}}^{I}\right) - E\left(\chi_{\text{child}}^{J}\right) \right|^{2}$$
$$V_{\text{pc}}^{IJ} = \left| \left\langle \chi_{\text{parent}}^{I} \left| V^{IJ}(\mathbf{R}) \right| \chi_{\text{child}}^{J} \right\rangle_{\mathbf{R}} \right|. \tag{31}$$

$\lambda$ is a parameter that drives the optimization iteratively. Minimizing (30) is equivalent to jointly minimizing the energy difference and maximizing the coupling between parent and child basis functions. For a fixed value of $\lambda$, minimization pushes the energy gap toward zero while maximizing the coupling as a function of $\mathbf{R}_{child}^{J}$ and $\mathbf{P}_{child}^{J}$. Sequentially increasing $\lambda$ steadily raises the penalty for energy non-conservation, while tracking changes in the coupling maximum as smoothly as possible. Each minimization cycle is performed with standard conjugate gradient techniques.

As illustrated in the examples of section IV, spawning optimally not only improves numerical efficiency by requiring fewer basis functions for branching ratio convergence, also provides insight into the physical character of nonadiabatic transitions. It reduces to simpler approaches such as the momentum-jump in appropriate limits, but remains exact for cases in which both position and momentum require adjustment (*e.g.*, tunneling phenomena).

## IV. RESULTS

The following test simulations define the initial wavefunction as a single multidimensional Gaussian, referenced as the target wavefunction. Propagation is performed on diabatic potential energy surfaces, although sometimes adiabats will be plotted for illustration purposes. The complex amplitude of each trajectory in the bundle is initialized by projecting the target wavefunction onto the individual basis functions,

$$c_{k}^{I}(0) = \sum_{n=1}^{N_{I}(0)} \left(S^{-1}\right)_{mn}^{II} \left\langle \chi_{n}^{I}(t=0) \middle| \Psi_{t=0}^{\text{target}} \right\rangle. \tag{32}$$

Using (32) to define the initial amplitudes guarantees that the initial FMS wavefunction has maximal overlap with the target.

### A. One-dimensional avoided crossing model

Results will be reported for three different spawning methods. Strict p-jump refers to spawning with pure momentum adjustment along the nonadiabatic coupling vector, similar to the surface hopping algorithm. Standard spawning is of the form most often implemented in previous FMS and AIMS simulations, where a steepest descent quench to the energy shell obviates frustrated spawns. The third method, optimal spawning, is the method introduced in III.C.3.

The first test system is a one-dimensional avoided crossing model, utilized often by Tully [22,42] and others.[43] The diabatic potential matrix is given by

$$V_{11}(X) = \begin{cases} A(1-\exp(-BX)) + A, & x > 0 \\ -A(1-\exp(BX)) + A, & x < 0 \end{cases}$$

$$V_{22}(X) = \begin{cases} -A(1-\exp(-BX)) + A, & x > 0 \\ A(1-\exp(BX)) + A, & x < 0 \end{cases} \quad (33)$$

$$V_{12}(X) = C\exp(-DX^2)$$

and parameters are assigned as in previous literature: $A = 0.01$, $B = 1.6$, $C = 0.005$, and $D = 1$. The corresponding adiabatic potential energy surfaces can be obtained by diagonalizing the two by two matrix defined by (33).

The system is prepared on the diabatic state 1 (with lower energy for $x < 0$), outside of the coupling region. The diabatic transmission coefficient will vary as a function of initial momentum, and its values for initial momentum $K_{initial}$ are plotted in the upper panel of Figure 1. For $K_{initial}$<4.5 a.u., a classical particle cannot surmount the (adiabatic) ground state potential barrier and will be reflected completely. For $K_{initial}$<8.9 a.u., the initial kinetic energy is below the asymptotic energy of the upper potential curve. If $K_{initial}$ falls in the range 7.7-8.9 a.u., the particle has enough energy to become trapped temporarily in the well of the adiabatic excited state. We refer the reader to the lower panel of Figure 1 for more detailed illustration of ground and excited electronic states. For convenient reference, the *y*-axis of the lower panel of Figure 1 corresponds to position, and the *x*-axis is $K_{initial}$ instead of energy.

In Figure 1, numerically exact quantum results obtained with the fast Fourier transform (FFT) method[44,45] are shown with the red curve. The initial wavepacket for the FFT method is a Gaussian placed at the same phase space point as that of the initial wavefunction for FMS. The Gaussian width is chosen to be $\sigma = \sqrt{20}/K_{initial}$, corresponding to an energy spread that is 10% of the quantum mechanical total. The same width is used to define the initial FMS wavefunction. Atomic masses are 2000 a.u., comparable to that of a hydrogen atom. Quantum effects, especially tunneling, are expected to influence transmission coefficients under these conditions.

All of the spawning methods compare well with numerically exact results at large initial kinetic energy. Agreement for large $K_{initial}$ values is not surprising, because trajectory methods are known to be effective for quantum systems in classical regimes. As shown in the figure, various spawning methods give maximum transmission coefficients around $K_{initial}$=7.7 a.u. When initial momentum drops below this value, classically there is no transmission between the two electronic states. If initial momentum is smaller than 4.5 a.u., the classical *reflection* coefficient should be 100%. Due to tunneling, however, transmission is nonzero even for kinetic energy < 4.5 a.u., and the transmission coefficient decreases smoothly rather than discontinuously as a function of energy. Transmission for spawning with strict momentum jump decreases sharply as a function of energy, while optimal spawning matches more closely the numerically exact results than does standard spawning. Both standard and optimal spawning give finite transmission coefficients for initial momenta smaller than 4.5 a.u.

A more detailed study of FMS convergence is shown in Figure 2, for initial momentum $K_{initial}$=5. As illustrated in the inset, this choice corresponds to the low energy case with non-negligible quantum effects. The spawning method with strict momentum jump converges to a transmitted population much lower than with FFT. Standard spawning converges reasonably well, while optimal spawning yields the most accurate result. It is also worthwhile to note that the final total number of trajectories is roughly equal regardless of spawning method. The three methods converge to the same degree of accuracy in the high-energy case of $K_{initial}$=15, as shown in Figure 3. This is not surprising given that, in general, even semiclassical methods are expected to provide good approximations to quantum mechanics at high-energy.[22,46]

## B. Two-dimensional conical intersection model

We now examine the performance of the various spawning procedures with a two-dimensional, two-state conical intersection model. As discussed by Ferretti, etc.,[47] this model has been used to describe collinear reaction of triatomic *ABA*, and provides a useful testing bed for study and comparison of nonadiabatic simulation schemes. There are two stretching coordinates, $R_1$ and $R_2$, and two diabatic electronic states (hereafter referred to as the state 1 and 2) in the electronic Hamiltonian. It is intuitively and numerically advantageous to change from $R_1$ and $R_2$ to symmetric $X = \frac{R_1 + R_2}{2}$ and antisymmetric $Y = \frac{R_1 - R_2}{2}$ stretch coordinates, enabling the Hamiltonian to be rewritten:

$$V_{11}(X,Y) = \frac{1}{2}k_x(X - X_1)^2 + \frac{1}{2}k_y Y^2 \tag{34}$$

$$V_{22}(X,Y) = \frac{1}{2}k_x(X - X_2)^2 + \frac{1}{2}k_y Y^2 + \Delta$$

$$V_{12}(X,Y) = \gamma Y e^{-\alpha(X - X_3)^2 - \beta Y^2}$$

The minima of the parabolic surfaces are located at $X_1 = 4, X_2 = 3$. The interstate coupling is controlled by parameter $\gamma$. The parameters in the Hamiltonian are assigned as follows: $k_x = 0.01$, $k_y = 0.1$, $\Delta = 0.01$, $\alpha = 3$, $\beta = 1.5$, $\gamma = 0.01$, and $X_3 = 3$, such that the conical intersection coincides with the excited state minimum. The potentials are illustrated in Figure 4.

Due to the nonadiabatic coupling, an incoming wavepacket traveling along the low-frequency, symmetric coordinate $X$ will remain quasistationary along the antisymmetric $Y$ coordinate. Positioned initially on the diabatic state 1, the total simulation time corresponds roughly to one half-period along the $X$ direction. Each trajectory is allowed to spawn in the nonadiabatic coupling region. The phase space location of the centroid of each parent trajectory is labeled $\left(\overline{\mathbf{R}}_{parent}^I, \overline{\mathbf{P}}_{parent}^I\right)$, and that of child trajectory is $\left(\overline{\mathbf{R}}_{child}^J, \overline{\mathbf{P}}_{child}^J\right)$. The difference between the various spawning methods lies in the assignment of $\left(\overline{\mathbf{R}}_{child}^J, \overline{\mathbf{P}}_{child}^J\right)$. Ultimately, the efficiency and robustness of assigning phase space locations to newly spawned basis functions should be reflected in the population transfer between the different electronic states.

In Figure 5, the wavepacket is initially placed at (*X=7, Y=0*) on the diabatic state 1, as shown in the inset. As the wavepacket evolves, population begins to transfer from state 1 to 2. Total population on state 2 is plotted as a function of time for initial basis sets of varying size. FMS results for 4 initial trajectories are shown in blue, those for 8 are shown in green, and numerically exact FFT[45,48] results are given in red. The accuracy of predicted branching ratios increases as one moves from strict momentum-jumps (solid line), to standard spawns (dashed line), to optimal spawns (dotted line). Regardless of spawning algorithm or treatment of frustrated spawns, FMS converges to exact quantum dynamics for large enough basis sets. Increasing the number of initial trajectories to eight (green lines) mutes differences among the various spawning methods.

Results for a similar test, but with lower initial energy, are presented in Figure 6. The system is started at $\left(x=5.2, y=0, P_x=0, P_y=0\right)$ on state 1. Note that this energy is only slightly above that of the conical intersection (cf. (34)), as shown in the inset in the upper left corner of Figure 6. The low energy case is challenging numerically, because most trajectories will not spawn due to insufficient kinetic energy near the intersection point; this is true especially for the strict p-jump approach. On the other hand, we expect optimal spawning to perform much better relative to other spawning methods when using small basis sets at low energy. As shown in Figure 6, for only 4 initial trajectories, optimal spawning (blue dotted line) captures both the peak and tail of the excited state population much better than the pure momentum-jump (blue solid line) or standard spawning (blue dashed line) methods. For detailed comparison of the smaller initial basis set, the peak region has been enlarged in the upper right inset. When the number of initial trajectories is increased to 8, optimal spawning (green dotted line) gives very accurate results, while population curves for strict p-jump (green solid line) and standard spawning (green dashed line) retain relatively large, spurious oscillations. These results illustrate that efficient capture of spawning events is critical to accurate population transfer, especially in the low energy case when the system has just enough energy to jump from the ground to the excited state.

In order to make more detailed comparison of the various spawning methods, we isolate a single spawning event around 3000 a.u. in Figure 7. The phase space location of the parent trajectory centroid is marked by a red square. At each point in coordinate space, the momentum is varied to check for kinetic energy sufficient to boost the parent trajectory from the ground to the excited state. The region of coordinate space with insufficient kinetic energy is shaded grey. Any spawn from the parent to a point in this grey area would be frustrated. Standard spawning (marked with the blue triangle) and pure p-jump methods yield child trajectory centroids located at the same coordinates as their parents. By definition, no coordinate shift accompanies a pure momentum jump; therefore, if the parent trajectory falls in the energetically "forbidden" region, it will lead to a frustrated spawn. When the standard spawning method is used, on the other hand, both coordinates and momenta may vary. A pure momentum adjustment along the nonadiabatic coupling vector direction is performed first. If the kinetic energy is insufficient, a descent quench perpendicular to the energy shell is followed by variation in coordinate space only. However, for the spawning event shown in Figure 7, the coordinate space location of the parent is the local potential minimum unless the momentum is also allowed to vary. Consequently, neither the pure momentum jump nor the standard spawning method is able to place the child outside the energetically forbidden region. By allowing both momentum and coordinate space variations, optimal spawning places the child trajectory at a phase space point (marked by the purple, solid circle) where energy is conserved and coupling is maximial. The contours in Figure 7 separate regions differing in coupling magnitude. The regions themselves are distinguished by varying shades of blue: the deeper the blue, the larger the coupling. The tuning parameter $\lambda$ in Eq. (30) is increased sequentially to drive the optimization of the child centroid, and a typical optimization path is indicated with the red curve. For illustration purposes, an unnecessarily small multiplicative step size of 1.1 has been chosen for $\lambda$ in Figure 7, so as to clarify its growth from one optimization cycle to the next. For standard applications, however, much larger steps in the range 2 to 10 are obtainable without sacrificing numerical accuracy. The numerical results reported in previous figures were generated with $\lambda$ increased by 1.01 times in every step.

    Optimal spawning allows newly spawned basis functions to be located further from their parent trajectory than the alternative spawning procedures, if additional displacement is required for energy conservation in frustrated spawns. This may be helpful in cases where a new trajectory needs to jump a large distance in coordinate

space, such as when tunneling through a barrier. Additional details, however, are beyond the scope of the current paper.

Nuclear wavefunction densities yielded by the various spawning methods are compared with numerically exact FFT in Figure 8. Snapshots are provided at the same time as that chosen in Figure 7. Because FMS with pure momentum jumps gives rise to densities very similar to those of standard spawning, only optimal and standard spawning methods are displayed. The salient feature of the optimal spawning density is its node along the *Y*-axis, in agreement with previous findings.[47] As pointed out by Ferreti, the presence of the node at *Y*=0 is a manifestation of Berry's phase[49], and cannot be reproduced by standard classical trajectory methods[47]. FMS with optimal spawning correctly treats nonadiabatic transitions near conical intersections, and is therefore able to reproduce even the effects of geometric phase.

## V. Conclusions

Optimal spawning introduces a rigorous procedure for assigning phase space points to spawned trajectories, subject only to the constraint of energy conservation. If a spawning event is not frustrated, optimal spawning leads to greater maximal coupling between the parent and child basis functions than alternative spawning procedures, which implies more efficient population transfer between the two electronic states. If a spawn is frustrated, it does not have to be abandoned as in the case of surface hopping. Frustrated spawns still are accepted, and the accuracy of the results is guaranteed by solution of the Schrödinger equation for the complex amplitudes. Optimal spawning improves upon previous implementations in its heightened promotion of population transfer in individual nonadiatic events. Its advantages increase with system dimensionality, since there are in general multiple choices of the new child basis function subject to the requirement of energy conservation and maximal coupling $V_{pc}^{IJ}$. Optimal spawning therefore provides a sophisticated and robust way to allow any phase space point to be chosen statistically for the newly spawned basis function in the nonadiabatic coupling region.

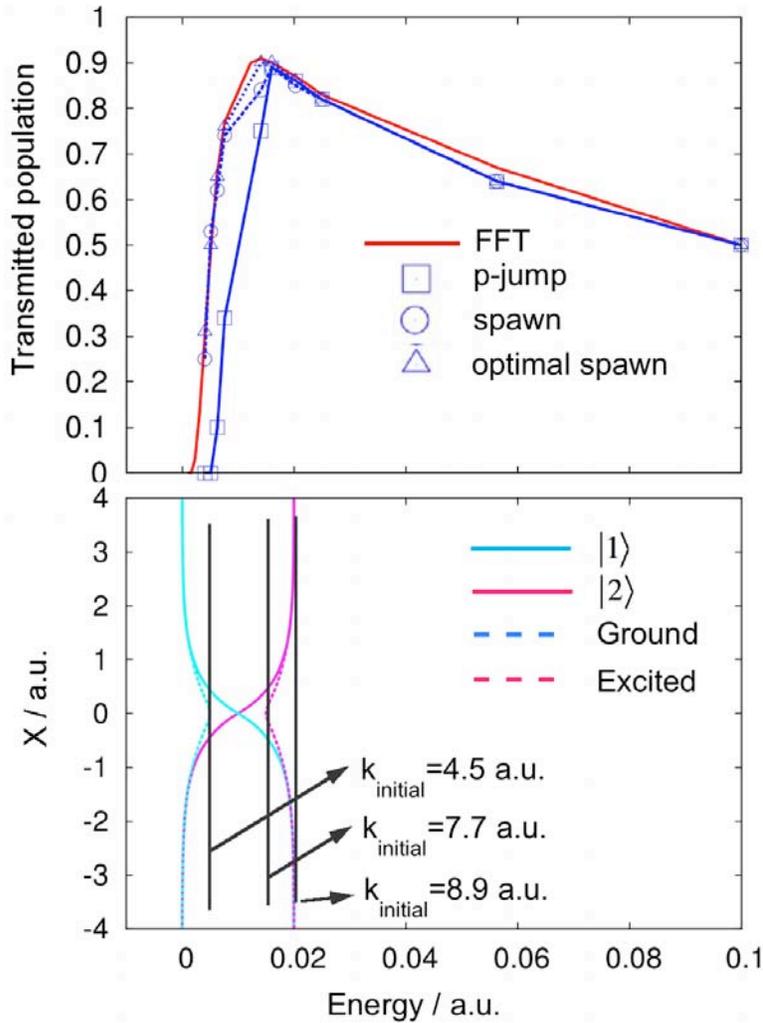

**Figure 1:** Diabatic transmission coefficients for the one-dimensional avoided crossing model. Upper panel: transmission coefficient vs. initial momentum $K_{initial}$. Numerically exact FFT results are shown in red, and FMS results using strict momentum-jumps, standard spawning and optimal spawning are depicted with blue squares, circles, and triangles, respectively. Lower panel: x-axis is $K_{initial}$, on the same scale as that of the upper panel. The two diabats for state 1 and 2 of the one-dimensional avoided crossing model are shown with green and purple solid lines, and ground and excited adiabats with green and purple dashed lines.

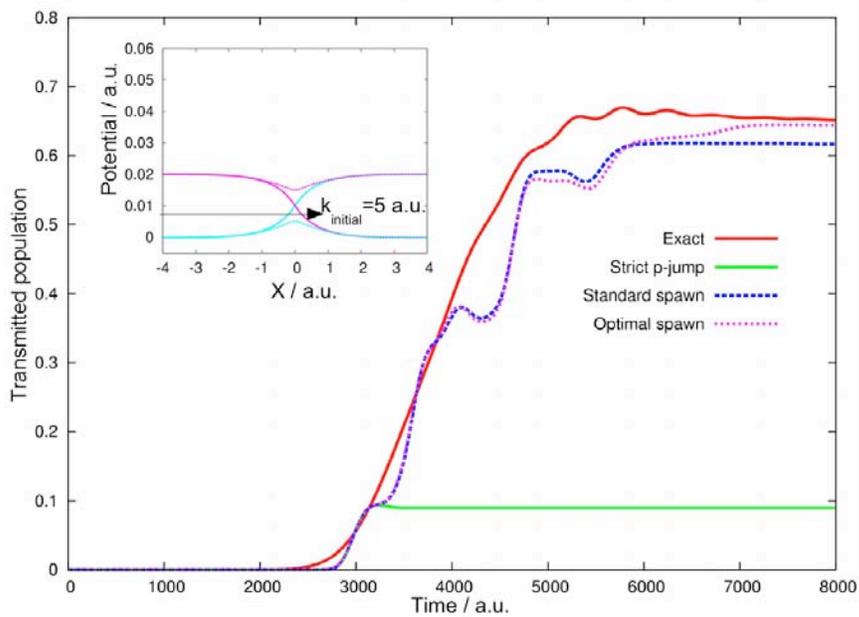

**Figure 2:** Transmitted population from diabatic state 1 to 2, as a function of time. The initial momentum of the wavepacket is $K_{initial}$=5, as shown in the inset. Red solid lines show exact FFT results. FMS using strict momentum jumps, standard spawning, and optimal spawning technique are shown with blue solid, dashed, and dotted lines respectively.

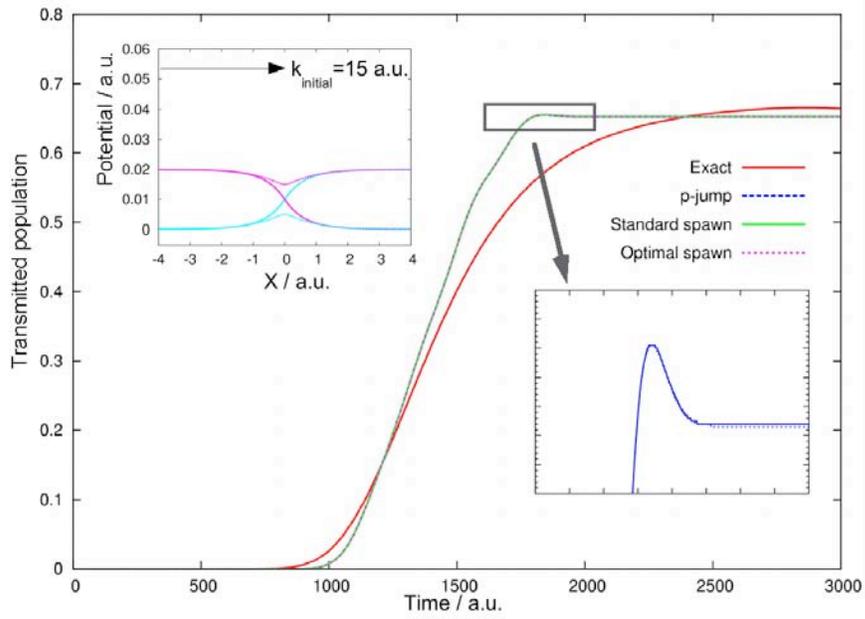

**Figure 3:** Transmitted population from diabatic state 1 to 2, for initial momentum $K_{initial}=15$.

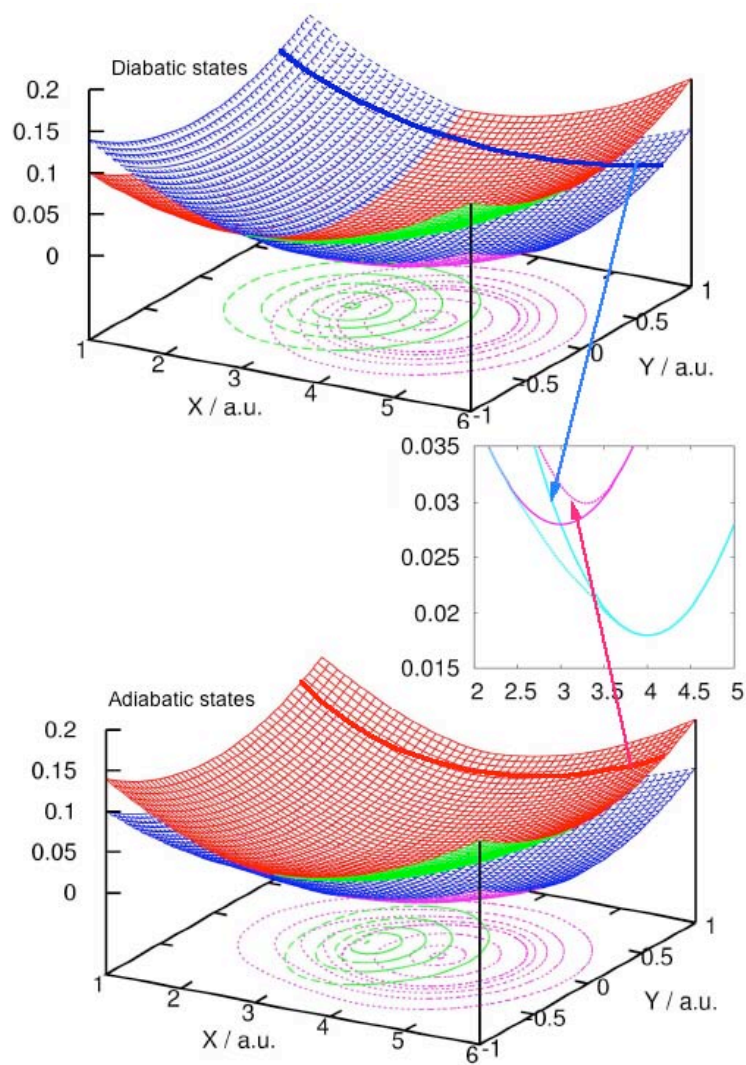

**Figure 4:** Diabatic and adiabatic potential energy surfaces for the two-dimensional, two-state Persico model. The one-dimensional cut (center) is taken at *Y=0.6*, where the nonadiabatic coupling is maximum.

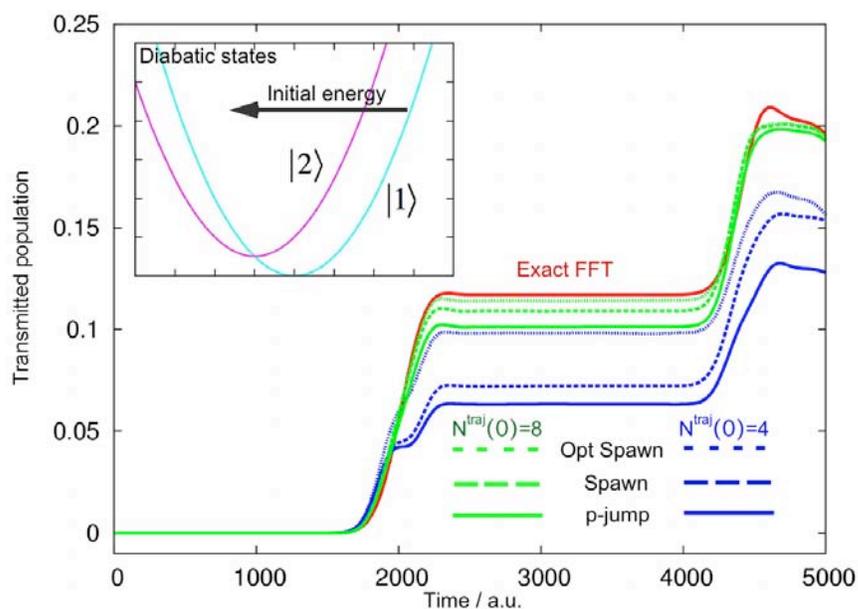

**Figure 5:** Excited state population as function of time. The initial wavepacket is placed at *X=7, Y=0* on the diabatic state 1. Exact FFT results are given in red. FMS results for 4 initial trajectories are shown in green, and for 8 initial trajectories are shown in red. For both sets of initial conditions, solid lines are used for FMS with strict momentum jumps, dashed lines for standard spawning, and dotted lines for optimal spawning.

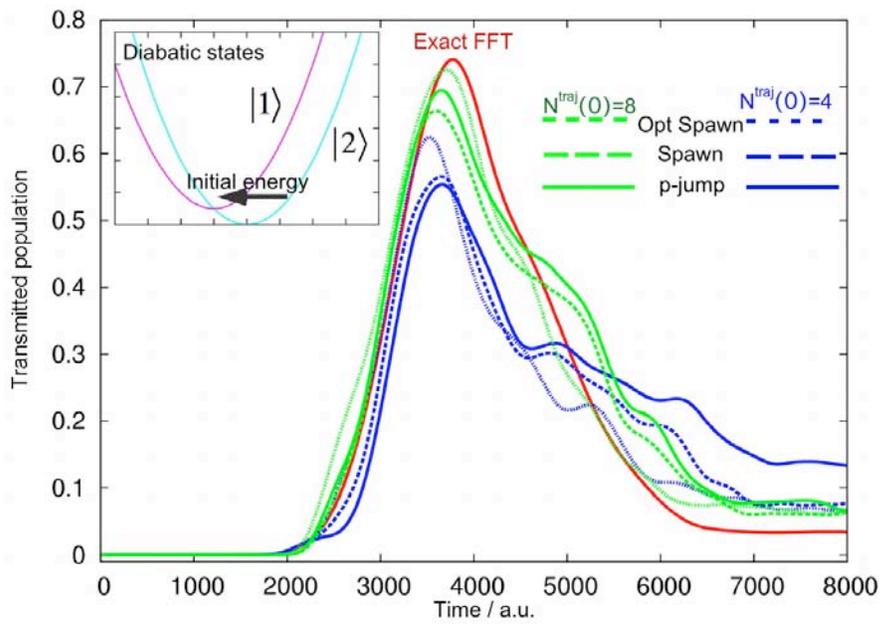

**Figure 6:** Excited state population as a function of time, for an initial wavepacket placed at *X=5.2, Y=0* on the diabatic state 1. Notations are the same as those used in Figure 5.

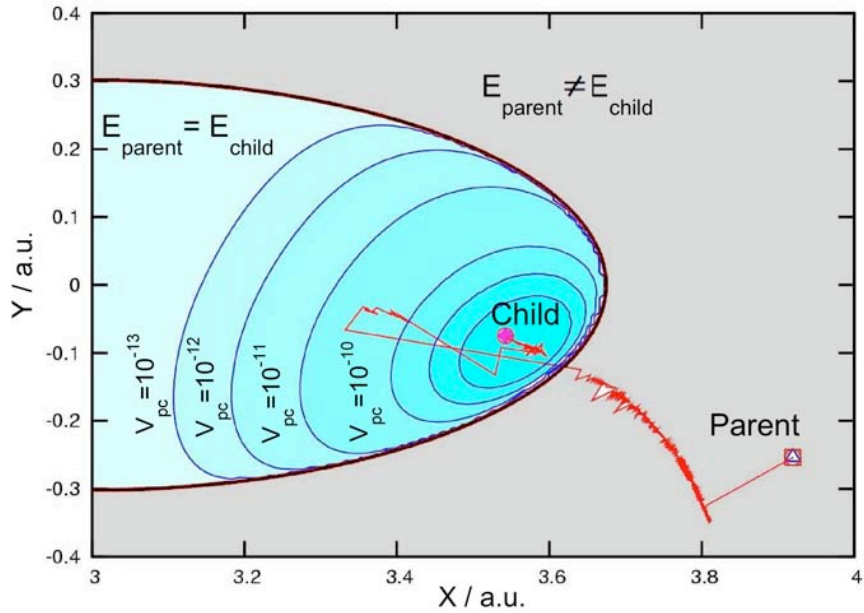

**Figure 7:** Initial conditions for a child trajectory using various spawning techniques. The centroid of the parent basis function is marked with a red square. For this particular spawn, strict momentum jump or standard spawning gives initial conditions identical to that of the parent. Optimal spawning moves the child trajectory to the phase space point marked with the red circle. The grey area is the energetically forbidden region where with $E_{parent} < E_{child}$. In the energetically allowed region, contour lines distinguish magnitudes of the coupling $V_{pc}$ between parent and child basis functions. For illustration purposes, very conservative thresholds have been employed and the optimization requires about 100 steps. Typically a much larger step in $\lambda$ would be used, leading to many less iteration steps.

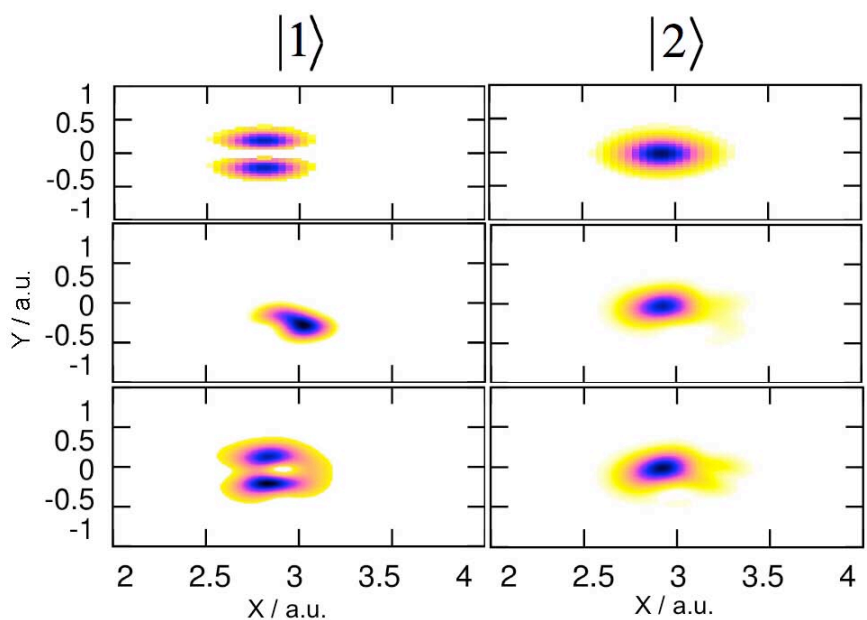

**Figure 8:** The shape of wavepackets in two-dimensional coordinates are plotted at the time just after nonadiabatic coupling starts to cause population transfer from ground to excited adiabatic states of the Persico model. Subfigures on the left show the wavepackets on the ground state, and those on the right for the excited state. FFT is compared against standard and optimal spawning algorithms. The wavepacket generated with FMS using strict momentum jump is very similar to that with standard spawning, and is thus not shown in this figure.